\documentclass[11pt]{article}
\usepackage{amsmath}
\usepackage[]{graphicx}
\usepackage[]{cite}

\newcommand{\dd}{\mathrm{d}}
\newcommand{\sv}{\mathrm{v}}
\newcommand{\tu}{\tilde{u}}
\newcounter{myitemCOUNTER} 
\newcommand{\myitem}[1]{%
\refstepcounter{myitemCOUNTER} \item[\themyitemCOUNTER.] \label{#1}}
\numberwithin{equation}{section}  

\newcommand{\bmath}{\begin{displaymath}}
\newcommand{\emath}{\end{displaymath}}
\newcommand{\beq}{\begin{equation}}
\newcommand{\eeq}{\end{equation}}
\newcommand{\beqa}{\begin{eqnarray}}
\newcommand{\eeqa}{\end{eqnarray}}

\newcommand{\gtrsim}{\stackrel{>}{\sim}}

\begin{document}

\title{Self-sustained nonlinear waves in traffic flow}
\author{M.~R.~Flynn$^1\/,\;$ A.~R.~Kasimov$^2\/,\;$ J.-C.~Nave$^2\/,\;$
        R.~R.~Rosales$^2$, \and B.~Seibold$^2$. \vspace*{5mm} \\
        $^1$ Dept.~of Mechanical Engineering, Univ.~of Alberta, \\
             Edmonton, AB, T6G 2G8 Canada \vspace*{5mm} \\
        $^2$ Dept.~of Mathematics, Massachusetts Inst.~of Technology, \\
             77 Massachusetts Avenue, Cambridge, MA 02139, USA} 

\date{\today}

\maketitle

\begin{abstract}
In analogy to gas-dynamical detonation waves, which consist of a shock
%
with an attached exothermic reaction zone,
we consider herein nonlinear traveling
wave solutions, termed ``jamitons,'' to the hyperbolic (``inviscid'')
continuum traffic equations. Generic existence criteria are examined in
the context of the Lax entropy conditions. Our analysis naturally
precludes traveling wave solutions for which the shocks travel downstream
more rapidly than individual vehicles. Consistent with recent experimental
observations from a periodic roadway (Sugiyama \emph{et al.~New Journal
of Physics}, \textbf{10}, 2008), our numerical calculations show that,
under appropriate road conditions, jamitons are attracting solutions,
with the time evolution of the system converging towards a
jamiton-dominated configuration. Jamitons are characterized by a sharp
increase in density over a relatively compact section of the roadway.
Applications of our analysis to traffic modeling and control are examined
by way of a detailed example.  

\end{abstract}

%
\section{Introduction and problem formulation}
\label{sec:ipf}
The economic costs in terms of lost productivity, atmospheric pollution
and vehicular collisions associated with traffic jams are substantial both
in developed and developing nations. As such, the discipline of traffic
science has expanded significantly in recent decades, particularly from
the point of view of theoretical modeling \cite{helbing2001}. Borrowing
terminology applied in Payne \cite{payne1979} and elsewhere, three generic
categories describe the approaches considered in most previous analyses.
``Microscopic'' models, such as ``follow the leader'' studies
\cite{pipes1953} or ``optimal velocity'' studies \cite{newell1961}
consider the individual (i.e.~Lagrangian) response of a driver to his or
her neighbors, in particular, the vehicle immediately ahead.
``Mesoscopic''
or ``gas-kinetic macroscopic'' analyses, such as the examinations of
Phillips \cite{phillips1979} and Helbing \cite{helbing2008a} take a
statistical mechanics approach in which vehicle interactions are modeled
using ideas familiar from kinetic theory. Finally ``macroscopic'' studies
\cite{payne1979, lighthill_whitham1955, richards1956, kerner_konhauser1993,
kerner_konhauser1994, aw_rascle2000, helbing2008b} model traffic flow
using conservation laws and a suitable adaptation of
the methods of continuum mechanics \cite{whitham1974, leveque1992},
which yields governing equations similar to those from fluid mechanics.
It is this latter category of analysis that is of interest here.

Treating the traffic flow as a continuum, we begin by considering a one
dimensional Payne-Whitham model with periodic boundary conditions, i.e.
vehicles on a circular track of length $0 < \lambda < \infty\/$
%
%
\cite{sugiyama_fukui_kikuchi_hasebe_nakayama_nishinari_tadaki_yukawa2008}.
The governing equations for mass and momentum are then
(e.g.~Kerner \& Konh{\" a}user \cite{kerner_konhauser1994})
\begin{eqnarray}
  \rho_t + (\rho\,u)_x & = & 0\/,  \label{eq:PWmass1}\\
  u_t+u\,u_x+\frac{1}{\rho}\,p_x & = & \frac{1}{\tau}\,
  \left(\tu-u\right)\/, \label{eq:PWmom1}
\end{eqnarray}
where the subscripts indicate differentiation, $\tau\/$ is a relaxation
time-scale, $u\/$ is the traffic speed, and $\rho\/$ is the traffic
density --- with units of vehicles/length. The traffic pressure, $p\/$,
which incorporates the effects of the ``preventive'' driving needed to
compensate for the time delay $\tau\/$, is typically assumed to be an
increasing function of the density only, i.e.~$p=p(\rho)\/$
\cite{aw_rascle2000}. Here, in order to have a well-behaved theoretical
formulation in the presence of shock waves \cite{evans1998}, we shall
assume that $p\/$ is a convex function of the specific volume
$\sv = 1/\rho\/$ (road length per vehicle). This implies that
$\dd\/p/\dd\/\sv < 0\/$ and $\dd^2\/p/\dd\/\sv^2 > 0\/$, which holds for
the functions typically assigned to $p\/$ in macroscopic models. Finally,
$\tu=\tu(\rho)\/$ gives, for a particular traffic density, the desired or
equilibrium speed to which the drivers try to adjust. The precise
functional form of $\tu\/$ is, to a certain degree, rather arbitrary and
indeed several variants have been proposed \cite{helbing2001}. Generally,
$\tu\/$ is a decreasing one-to-one function of the density, with
$0<\tu(0) = \tu_0<\infty\/$ and $\tu(\rho_M)=0\/$, where:
\begin{itemize}
 %
 \item
 $\rho_M\/$ denotes the maximum density, at which the vehicles are nearly
 ``bumper-to-bumper'' --- thus $\ell = \rho_M^{-1}\/$ is the ``effective''
 (uniform) vehicle length.
 %
 \item
 $\tu_0\/$ is the drivers' desired speed of travel on an otherwise
 empty road.
 %
 %
\end{itemize} \vspace*{-1mm}
In this paper, we restrict ourselves to the representative form
$\tu = \tu_0\,(1-\rho/\rho_M)^n\/$, where $n\/$ is ``close'' to $1\/$.
%
%
%
We defer the detailed treatment of the exact conditions on $\tu\/$ that
guarantee the existence of self-sustained nonlinear traveling waves in
traffic (termed ``jamitons'' herein) to a later publication.

Ubiquitous attributes of the solutions to continuum traffic models are
stable shock-like features (see e.g.~Kerner \& Konh{\" a}user
\cite{kerner_konhauser1993} and Aw \& Rascle \cite{aw_rascle2000}). In
analyzing such structures, a dissipative term proportional to $u_{xx}\/$,
analogous to the viscous term in the Navier-Stokes equations, is often
added to the right-hand side of the momentum equation (\ref{eq:PWmom1})
in order to ``smear out'' discontinuities \cite{helbing2001}. However,
the physical rationale for this term is ambiguous and the proper
functional form is therefore subject to debate. Solutions, such
as those obtained by Kerner \& Konh{\" a}user \cite{kerner_konhauser1993},
whose dynamics are non-trivially influenced by viscous dissipation must
therefore be interpreted with care.
Herein, an alternative line of inquiry is proposed: we seek
self-sustained traveling wave solutions to the ``inviscid'' equations
(\ref{eq:PWmass1} -- \ref{eq:PWmom1}) on a periodic domain, where shocks
are modeled by discontinuities, as in the standard theory of shocks for
hyperbolic conservations laws \cite{evans1998}. As we demonstrate below,
not only do such nonlinear traveling waves exist, but they have a
structure similar to that of the self-sustained detonation waves in the
Zel'dovich-von~Neumann-Doering (ZND) theory \cite{fickett_davis1979}.
According to the ZND description, detonation waves are modeled as shock
%
%
waves with an attached exothermic reaction zone.
In a self-sustained detonation wave, the flow downstream of the shock is
subsonic relative to the shock, but accelerating to become sonic
at some distance away from the shock. 
%
%
%
Hence, the flow behind a self-sustained
detonation wave can be ``transonic,'' i.e.~it may undergo
a transition from subsonic to supersonic.
The existence of the sonic point, the location where the flow speed relative to the shock equals the local sound speed, is the key feature in the ZND theory
that allows one to solve for the speed and structure of the detonation
wave.
Its existence also means that the shock wave cannot be influenced by
smooth disturbances from the flow further downstream so that the shock
wave becomes self-sustained and independent of external driving
mechanisms. 
Hence the sonic point is an ``acoustic'' information or event horizon \cite{StewartKasimovSIAP05}.

For the nonlinear traffic waves to be discussed herein,
this means that their formation, due to small initial perturbations,
is analogous to the ignition and detonation that can occur in a
meta-stable explosive medium.
Although this analogy has not, to our knowledge, been reported
in the traffic literature, the physical and mathematical similarities
between detonation waves and hydraulic jumps, described by equations
similar to (\ref{eq:PWmass1}) and (\ref{eq:PWmom1}), were recently pointed
out by Kasimov \cite{kasimov2008}. (The analogy between hydraulic jumps
and {\it inert} gas-dynamic shocks was recognized much earlier -- see
e.g.~Gilmore {\it et al.}~\cite{gilmore_plesset_crossley1950} and Stoker
\cite{stoker1957}.) As with Kasimov's analysis, our aim is to herein
exploit such commonalities to gain additional understanding into the
dynamics of traffic flows, in particular, the traffic jams
%
%
%
that appear in the absence of bottlenecks and for no apparent reason. From
this vantage point, novel insights are discerned over and above those that
can be realized from the solution of a Riemann problem
\cite{aw_rascle2000} or from the linear stability analysis of uniform base
states (see e.g.~Appendix \ref{sec:appC}). Indeed, when such linear
instabilities are present initially, our extensive numerical experiments
suggest that the resulting ``phantom jams'' (see Helbing
\cite{helbing2001} and the many references therein) will ultimately
saturate as jamitons. This
observation provides a critical link between the initial and final
states, the latter of which can, under select conditions (see
e.g.~\S~\ref{sec:res}), be described analytically. Moreover, as we plan
to illustrate in forthcoming publications, such self-sustained traffic
shocks are also expected on non-periodic roads. Thus the model results
presented below can be readily generalized beyond the mathematically
convenient case of a closed circuit. 

The rest of the paper is organized as follows: in \S~\ref{sec:PW}, we
outline the basic requirements for (\ref{eq:PWmass1}) and
(\ref{eq:PWmom1}) to exhibit traveling wave solutions. To demonstrate the
generality of this analysis, we consider in \S~\ref{sec:AR} modified
forms for the momentum equation (developed by Aw \& Rascle
\cite{aw_rascle2000} and Helbing \cite{helbing2008b}). From this different
starting point, the salient details of \S~\ref{sec:PW} shall be
reproduced. The analysis is further generalized in \S~\ref{sec:pp}, which
considers a phase plane investigation of the governing equations from
\S~\ref{sec:PW} and \S~\ref{sec:AR}. A particular example is studied, both
theoretically and numerically, in \S~\ref{sec:example} in which
$\tu\/$ and $p\/$ are assigned particular functional forms. The
impact of our findings on safe roadway design is briefly discussed.
Conclusions are drawn in \S~\ref{sec:conc}. 
%
%
\section{Traveling wave solutions -- jamitons}
\label{sec:PW}
To determine periodic traveling wave solutions to the traffic flow
equations (\ref{eq:PWmass1}) and (\ref{eq:PWmom1}), we begin by making
the solution ansatz, $\rho=\rho\left(\eta\right)\/$ and
$u=u\left(\eta\right)\/$, where the self-similar variable $\eta\/$ is
defined by
\begin{equation} \label{eq:eta}
  \eta=\frac{x-s\/t}{\tau}\/.
\end{equation}
Here $s\/$ is the speed, either positive or negative, of the traveling
wave. Equation (\ref{eq:PWmass1}) then reduces to
\begin{equation} \label{eq:PWmass2}
  \left\{ \rho\left(u-s\right)\right\} _{\eta}=0,
  \quad \Longrightarrow \quad \rho=\frac{m}{u-s}\/,
\end{equation}
where the constant $m\/$ denotes the mass flux of vehicles in the wave
frame of reference. Substituting (\ref{eq:PWmass2}) into
(\ref{eq:PWmom1}), we obtain
\begin{equation} \label{eq:PWmom2}
  \frac{\dd\/u}{\dd\/\eta} = \frac{\left(u-s\right)\,
  \left(\tu-u\right)}{\left(u-s\right)^{2}-c^{2}}\/,
\end{equation}
where we interpret $\tu=\tu(\rho)\/$ as a function of $u\/$ via (\ref{eq:PWmass2}). Here
$c=(p_{\rho})^{1/2}>0\/$ is the ``sound speed,'' i.e.~the speed at which
infinitesimal perturbations move relative to the traffic flow.

Equation (\ref{eq:PWmom2}) is a first order ordinary differential equation
and therefore, barring pathological and unphysical choices for $c\/$ and
$\tu\/$, does not admit any smooth periodic solutions. Hence, the periodic
traveling wave(s) --- if they exist --- must consist of monotone solutions to
(\ref{eq:PWmom2}) that are connected by shocks. The simplest situation, as
reproduced in the physical experiment of Sugiyama \emph{et al}.~%
\cite{sugiyama_fukui_kikuchi_hasebe_nakayama_nishinari_tadaki_yukawa2008},
is one in which there is exactly one shock (with speed $s\/$) per period.
The case of multiple shocks per period is more complex. We briefly address
this situation in \S~\ref{sec:res}.

%
Before going into the details of the solution, we recall that shocks must
satisfy two sets of conditions to be admissible. First, they must satisfy
the Rankine-Hugoniot conditions \cite{aw_rascle2000,  whitham1974,  leveque1992},
which follow from the conservation of mass and momentum, and ensure that
shocks do not become sources or sinks of mass and/or momentum. For the
equations in (\ref{eq:PWmass1} -- \ref{eq:PWmom1}), the Rankine-Hugoniot
conditions take the form,
\begin{eqnarray}
  s\,[\rho]   & = & \parbox{0.8in}{$[\rho u]\/$    }
                    \mbox{(i.e.~conservation of mass),}
  \label{eq:RH1}\\
  s\,[\rho u] & = & \parbox{0.8in}{$[p+\rho u^2]\/$}
                    \mbox{(i.e.~conservation of momentum),}
  \label{eq:RH2}
\end{eqnarray}
where $s\/$ is the shock speed and the brackets $[\;]\/$ indicate the jump
in the enclosed variable across the shock discontinuity. These equations
relate the upstream and downstream conditions at the
shock. In particular, let the superscripts $+\/$ and $-\/$ denote the
states immediately downstream (right) and upstream (left) of the shock,
respectively. Then (\ref{eq:RH1}) is equivalent to
\begin{equation} \label{eq:RH1expanded}
  \rho^+\,(u^+-s) = m = \rho^-\,(u^--s)\/,
\end{equation}
where the constant $m\/$ is the mass flux across the shock. Of course,
for a shock embedded within a jamiton, this $m\/$ is the same as the
one in (\ref{eq:PWmass2}).

\smallskip
Second, shocks must satisfy the Lax ``entropy'' conditions
\cite{leveque1992, evans1998}, which enforce dynamical stability. In the
case of shocks in gas-dynamics, these conditions are equivalent to the
statement that the entropy of a fluid parcel increases as it goes through
the shock transition \textemdash{} hence the name. However, the existence
of a physical entropy is not necessary for their formulation: stability
considerations alone suffice. Furthermore, these conditions also guarantee
that the shock evolution is causal.

\begin{figure}
 \vspace*{-4cm} \hspace*{1.5cm}
 \includegraphics[width=10cm]{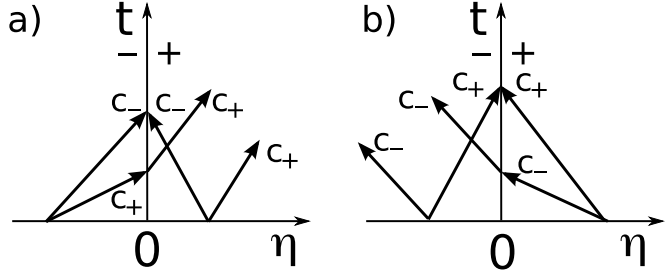}
 \noindent
 \vspace*{-6.5cm}
\caption{Characteristics on both sides of a left (a) and right (b) shock
   in the frame of the shock. The flow direction is from left-to-right in (a)
   and right-to-left in (b). The equations of the characteristics are
   $C_{\pm}:\;\dd\/\eta/\dd\/t = (u-s\pm c)/\tau\/$, where
   $\eta\/$ is defined by (\ref{eq:eta}), and $s\/$ is the shock speed.}
\label{fg:leftshockrightshock}
\end{figure}

For the particular system of equations in
(\ref{eq:PWmass1} -- \ref{eq:PWmom1}), the Lax entropy conditions ---
given below in (\ref{eq:Laxleft} -- \ref{eq:Laxright}) --- state that one
family of characteristics\footnote{The curves in space-time
   along which infinitesimal perturbations propagate. Namely:
   $\dd\/x/\dd\/t = u+c\/$ (``right'' characteristics),
   and
   $\dd\/x/\dd\/t = u-c\/$ (``left'' characteristics).}
must converge into the shock path in space-time, while the other family
must pass through it. Thus exactly two families of shocks are possible, as
illustrated in figure~\ref{fg:leftshockrightshock}: The left (respectively
right) shocks have the left (respectively right) characteristics
converging upon them. 
%
%
%
%
%

In the context of the periodic traveling waves with a single
shock per period, the above discussion implies that, in principle, two
cases are possible: jamitons containing a left shock wherein the mass flux, $m\/$, is positive -- see item \ref{itm:Laxleft} below, and jamitons containing a right shock wherein the mass flux, $m\/$, is negative -- see item \ref{itm:Laxright} below. However, as we argue after item~\ref{itm:Laxright},
%
%
only jamitons with $m > 0\/$ are mathematically consistent; self-sustained traveling waves carrying within them a right shock are not permitted. 
This is consistent with the experiment of Sugiyama {\it et al.}
\cite{sugiyama_fukui_kikuchi_hasebe_nakayama_nishinari_tadaki_yukawa2008},
as well as with one's everyday driving experience: it is situations where
individual vehicles overtake shocks (hence $m > 0\/$) that are observed in reality, rather than the converse. 
Thus whereas in second order traffic models information in the form of shock waves can travel downstream faster than individual vehicles \cite{daganzo1995}, the results in this paper show that this cannot happen in the form of a self-sustained traveling wave. Further to the analysis of
Aw \& Rascle \cite{aw_rascle2000} and Helbing \cite{helbing2008b}, this
observation lends additional support to the conclusion that second order
models are not \emph{ipso facto} flawed \textemdash{} see also Helbing
\cite{helbing2001}, \S~III.D.7 and the references therein.

Let us now consider in some detail the two scenarios that can, in principle, arise for traveling waves with a single shock per period.
%
%
%
\begin{itemize}
  \myitem{itm:Laxleft}
  The shock is the left shock, as in figure
  \ref{fg:leftshockrightshock}\,a, so that
  \begin{equation} \label{eq:Laxleft}
     \left(u-c\right)^{-} > s > \left(u-c\right)^{+}\/.
  \end{equation}
  %
  %
  In this case the mass flux must be positive, since
  $m = \rho^{-}\,\left(u^--s\right) > \rho^{-}\,c^{-} > 0\/$. Moreover,
  $\rho^{+}/\rho^{-} = \left(u-s\right)^{-}/\left(u-s\right)^{+} >
    c^{-}/c^{+}\/$, so that $\rho^{+}c^{+}>\rho^{-}c^{-}\/$. The traffic
  pressure $p\/$ is a convex function of $\sv=\rho^{-1}\/$ (see
  \S~\ref{sec:ipf}), hence $\rho\,c = (-\dd\/p/\dd\/\sv)^{1/2}\/$ is an
  increasing function of the traffic density. Thus
  $\rho^{+}\,c^{+} > \rho^{-}\,c^{-}$ implies that $\rho^{+}>\rho^{-}\/$. In
  other words, the shock is compressive: the traffic density increases
  as the vehicles pass through the shock, traveling from left to right. 
  Conversely, since $u=s+m/\rho\/$, it follows that $u^{-}>u^{+}\/$ and, consequently, vehicles decelerate as they overtake the shock. It should then
  be clear that these shocks have all the familiar properties of traffic jams.

  We conclude that for a traveling wave with a
  left shock, the continuous solution of (\ref{eq:eta} -- \ref{eq:PWmom2})
  must have a decreasing density, $\dd\/\rho/\dd\/\eta < 0\/$, and an
  increasing velocity, $\dd\/u/\dd \/\eta > 0\/$. This follows because the
  solution must be a monotone function of $\eta\/$, and must connect the
  post-shock state
  $\left(\rho^{+}\/,\,u^{+}\right)\/$ in one shock, to the pre-shock state
  $\left(\rho^{-}\/,\,u^{-}\right)\/$ in the subsequent shock across a period in
  $\eta\/$ --- say from $\eta = 0\/$ to  $\eta = \lambda\/$.
  \myitem{itm:Laxright}
  The shock is the right shock, as in figure \ref{fg:leftshockrightshock}\,b, so that
  \begin{equation} \label{eq:Laxright}
    \left(u+c\right)^{-} > s >\left(u+c\right)^{+}\/.
  \end{equation}
  The mass flux now is negative, since
  $m = \rho^{+}\,\left(u-s\right)^{+} < -\rho^{+}\,c^{+} < 0\/$. As
  in item~\ref{itm:Laxleft}, it is straightforward to show that $u^- > u^+\/$
  and $\rho^- > \rho^+\/$. 
  %
  %
  %
  %
  For a traveling wave with a right shock, the
  continuous and monotone solution of (\ref{eq:eta} -- \ref{eq:PWmom2})
  must have both the density and velocity increasing with $\eta\/$ (i.e.~$\dd\/\rho/\dd\/\eta > 0\/$ and $\dd\/u/\dd \/\eta > 0\/$), in order to connect $\left(\rho^{+}\/,\,u^{+}\right)\/$ to
  $\left(\rho^{-}\/,\,u^{-}\right)\/$ across a period in $\eta\/$.
  Notice that the shock is again compressive: the traffic density (in the vehicles' frame of reference) increases as vehicles pass through the shock. However, in this latter case, the shock overtakes the vehicles from behind, which accelerate as they pass through the shock transition. 
  This is a clearly counter-intuitive situation, not observed in real traffic \cite{daganzo1995}.

  %
  %
\end{itemize}
%
Fortunately, as we demonstrate next, traveling wave solutions with $m < 0\/$ are mathematically inconsistent, which obviates the need to consider them any further.
First, (\ref{eq:PWmass2}) is employed to rewrite (\ref{eq:PWmom2}) in the form
\begin{equation} 
\label{eq:PWmomRRR}
 \frac{\dd\/u}{\dd\/\eta} = m\,\rho\, G(\rho\/,\,s\/,\,m)\/,
\end{equation}
where 
\begin{equation} 
\label{eq:PWmomRRRG}
 G = \frac{\tu-u}{m^2 - \rho^2\,c^2}\/,
\end{equation}
$m^2 - \rho^2\,c^2 = \rho^2\,\left\{ (u-s)^2 - c^2\right\}\/$,
and $u = s + m/\rho\/$. Because $\mathrm{d} u/\mathrm{d} \eta>0\/$, a smooth
solution connecting $\left(\rho^{+}\/,\,u^{+}\right)\/$ to
$\left(\rho^{-}\/,\,u^{-}\right)\/$ 
requires $G < 0\/$. 
However, $m^2-\rho^2\,c^2 = \rho\,(u-s+c)\,(m-\rho\,c)\/$ and $m - \rho\,c < 0\/$, and it follows from (\ref{eq:Laxright}) that
$(m^2 - \rho^2\,c^2)^+ > 0 > (m^2 - \rho^2\,c^2)^-\/$. Thus $G < 0\/$ requires
$(\tu - u)^+ < 0 < (\tu - u)^-\/$, which is impossible since
$\tu(\rho) - u = \tilde{u}(\rho) - s - m/\rho\/$ is a strictly decreasing function of $\rho\/$: 
$\tilde{u}(\rho)\/$ decreases with increasing $\rho$ by assumption and $m < 0\/$.

%
%

\medskip
%
The difficulties documented in the previous paragraph are avoided for traveling waves with
%
%
$m>0$. In this case, we demand a solution of
(\ref{eq:PWmomRRR}) with $\dd\/u/\dd\/\eta > 0\/$ and
$\dd\/\rho/\dd\/\eta < 0\/$, connecting
$\left(\rho^{+}\/,\,u^{+}\right)\/$ to
$\left(\rho^{-}\/,\,u^{-}\right)\/$. 
This in turn requires $G>0\/$ for $\rho^- < \rho < \rho^+\/$.
The assumptions on $p = p(\rho)\/$ imply that $m^2-\rho^2\,c^2\/$ is a strictly decreasing function of $\rho\/$ (item~\ref{itm:Laxleft}). Since $m^2-\rho^2\,c^2 = \rho\,(u-s-c)\,(m+\rho\,c)\/$ with $m + \rho\,c > 0\/$, it follows from (\ref{eq:Laxleft}) that $(m^2 - \rho^2\,c^2)^- > 0 > (m^2 - \rho^2\,c^2)^+\/$.
We conclude therefore that $m^2-\rho^2\,c^2\/$ has a unique, multiplicity one, zero
at some $\rho_s$, with $\rho^- < \rho_s < \rho^+\/$. In order
for (\ref{eq:PWmomRRR}) 
to have a smooth solution with the desired properties, the numerator
$\tu - u = \tu(\rho) - s - m/\rho\/$ must be such that
\begin{equation} \label{eqn:rrr-2.11}
 \left. \begin{array}{rl}
   \mbox{(a)} & \tu - u \;\;\mbox{has a simple zero at}\;\;
                 \rho = \rho_s\/;\\
   \mbox{(b)} & \tu - u > 0 \;\;\;\mbox{for}\;\;\;
                 \rho^- \leq \rho < \rho_s\/; \rule{0mm}{4.5mm}\\
   \mbox{(c)} & \tu - u < 0 \;\;\;\mbox{for}\;\;\;
                 \rho_s\, < \rho \leq \rho^+\/. \rule{0mm}{4.5mm}
 \end{array} \right\}
\end{equation}
This then guarantees not only that the zero in the denominator of
(\ref{eq:PWmomRRR} -- \ref{eq:PWmomRRRG}) is cancelled by a zero of the same order in the numerator, but that the resulting regularized ordinary differential equation yields $\dd\/u/\dd\/\eta > 0\/$ everywhere, as needed. Indeed, $\rho_s$ is the traffic density at the sonic point, with corresponding flow speed $u_s = s + m/\rho_s\/$.

\begin{figure}
 \vspace*{-6.25cm} \hspace*{-0.5cm}
 \includegraphics[width=13cm]{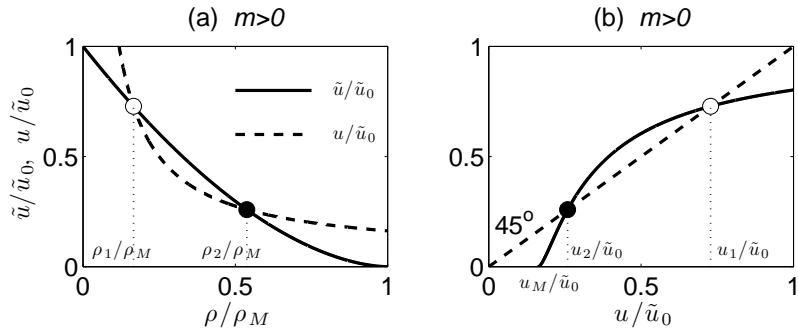}
 \noindent
 \vspace*{-9.cm}
\caption{Panel (a) shows typical profiles for $\tu/\tu_0\/$ (solid curve)
   and $u/\tu_0\/$ (dashed curve) as functions of $\rho/\rho_M\/$, where
   $u=s+m/\rho\/$. Equivalently, these profiles may be plotted against
   $u/\tu_0\/$ by employing the transformation (\ref{eq:PWmass2}), as
   shown in panel (b) --- the dashed curve is then just a line with unit
   slope.  The two curves will either not intersect at all, be tangent at
   a single point, or have two transversal intersections. The case of
   interest to us is the one with two transversal intersections, as
   depicted here:
   $u_2\/$ is the intersection with a smaller velocity and a larger
   density $\rho_2 = m/(u_2-s)\/$ --- solid circles, while the other
   intersection (open circles) defines $u_1\/$ and $\rho_1\/$.
   %
   It should be clear that, in order to satisfy the conditions in
   (\ref{eqn:rrr-2.11}), the sonic point must coincide with
   $(u_2\/,\,\rho_2)\/$. Finally, a physically meaningful solution
   requires $u > 0\/$ everywhere. Hence, $u_2 > 0\/$ is needed. As shown
   by panel (b), this condition is equivalent to the statement: when
   $\tu = 0\/$, that is $\rho = \rho_M\/$, the corresponding
   $u = u_M = s+m/\rho_M\/$ is positive.}
\label{fg:cartoon1}
\end{figure}

Figure~\ref{fg:cartoon1} illustrates the situation, with 
plots of $u\/$ and $\tilde{u}\/$ as functions of $\rho\/$ and $u\/$ for representative initial conditions $\rho^+\/$ and $u^+\/$. 
The shock speed, $s$, is here restricted by the inequalities $u^+ > s > (u - c)^+\/$. 
Clearly, the conditions in (\ref{eqn:rrr-2.11}) require that the sonic point values of the density, $\rho_s\/$, and velocity, $u_s\/$, coincide
with $\rho_2\/$ and $u_2\/$. The sonic condition therefore reads
\beq \label{eq:sonic0}
  u_2 = s+c\/(\rho_2)\/,
\eeq
from which the jamiton speed, $s\/$, can be determined. With $\rho_2\/$ and $u_2\/$ defined as functions of $(\rho^+\/,\,u^+\/,\,s)\/$ via figure~\ref{fg:cartoon1}, (\ref{eq:sonic0}) is an algebraic
equation that determines $s\/$ as a function of $(\rho^+\/,\,u^+)\/$. In general,  (\ref{eq:sonic0}) must be solved numerically, however, in \S~\ref{sec:example} we
provide an example where an analytic solution is possible. Finally, we
point out that {\it (i)} the restriction $u_M > 0\/$ must be imposed, where
$u_M=s+m/\rho_M\/$ is the speed corresponding to the maximum traffic density, $\rho_M$; {\it (ii)} the cases of $u_2 = u_1\/$, where $u_1$ is defined in figure \ref{fg:cartoon1}, or of no intersections between the solid and dashed curves of figure~\ref{fg:cartoon1} do not yield jamitons.
The methodology summarized above is reminiscent of the related analyses
in gas dynamics \cite{fickett_davis1979, StewartKasimovSIAP05},
%
%
shallow water theory
\cite{kasimov2008, dressler1949, balmforth_mandre2004}, astrophysical
accretion flow \cite{chakrabarti1990}, and Newtonian flow in elastic
tubes \cite{elad_kamm_shapiro1989}, where ordinary differential equations
similar to (\ref{eq:PWmom2}) are obtained. Indeed, (\ref{eq:sonic0}) is
the exact analog of the Chapman-Jouguet condition in detonation theory
\cite{fickett_davis1979}. 

Summarizing the above discussion, the following algorithm may be applied
to determine the jamiton structure:
\begin{itemize}
 \item[\emph{(i)}] \vspace*{-2mm}
 For a prescribed downstream state, $(\rho^+\/,\,u^+)\/$, the
 regularization condition (\ref{eq:sonic0}) specifies the permissible
 value(s) for the wave speed, $s\/$.
 %
 \item[\emph{(ii)}] \vspace*{-2mm}
 Once $s\/$ is determined, the state upstream of the shock,
 $(\rho^-\/,\,u^-)$, is computed using the Rankine-Hugoniot conditions
 (\ref{eq:RH1} -- \ref{eq:RH2}).
 \item[\emph{(iii)}] \vspace*{-2mm}
 Equation (\ref{eq:PWmom2}) is then integrated forward in $\eta\/$, from
 the initial condition $u=u^+\/$ up until $u=u^-\/$ is reached --- the
 traffic density, $\rho\/$, follows automatically from (\ref{eq:PWmass2}). This
 defines the period $\lambda\/$ of the traveling wave or, equivalently,
 the circumference of the periodic roadway.
 \item[\emph{(iv)}] \vspace*{-2mm}
 The total number of vehicles, $\mathcal{N}\/$, which remains fixed in
 time when there are no on-ramps or off-ramps, is then evaluated from
 \beq \label{eq:mathcalN1}
   \mathcal{N}=\int_0^{\lambda} \rho\,\dd\/x\/.
 \eeq
\end{itemize}
The jamitons thus obtained have the following properties:
{\it (i)} The traffic speed smoothly increases in the downstream direction
($\dd\/u/\dd\/\eta > 0\/$), except at the location of the shock, across
which there is an abrupt drop in $u\/$;
{\it (ii)}
The traffic density smoothly decreases in the downstream direction
($\dd\/\rho/\dd\/\eta < 0\/$), except at the location of the shock, across which there is an abrupt increase in $\rho\/$.
This is consistent with the experimental observations of
Sugiyama {\it et al.}~%
\cite{sugiyama_fukui_kikuchi_hasebe_nakayama_nishinari_tadaki_yukawa2008}.

The above algorithm provides a parameterization of the periodic jamitons
using $(\rho^+\/,\,u^+)\/$. Equivalent parameterizations, in terms of
$(\rho^-\/,\,u^-)\/$, are just as easy to produce. However, these are not
necessarily ideal parameterizations. For example, in order to predict what
jamiton configuration might arise from a given set of initial conditions,
on a given closed roadway,\footnote{Say to predict the patterns arising
   in the experiments by Sugiyama {\it et al.}
   \cite{sugiyama_fukui_kikuchi_hasebe_nakayama_nishinari_tadaki_yukawa2008}.}
a parameterization in terms of the roadway length, $\lambda\/$, and the
total number of vehicles, $\mathcal{N}\/$, would be more desirable. On
the other hand, for the example considered in \S~\ref{sec:example}, and
all the other case studies that we have examined to date, $\rho^-\/$ maps
in a one-to-one fashion to $\mathcal{N}\/$. Thus, by applying the above
algorithm, one can iteratively determine the unique traveling wave
solution corresponding to particular choices for the functions $p\/$ and
$\tu\/$, the parameters $\tau\/$ and $\lambda\/$, and the average traffic
density $\rho_{avg}=\mathcal{N}\lambda^{-1}\/$. 
%
%
\section{Alternative description of the momentum equation}
\label{sec:AR}
The existence of self-sustained shock solutions is not specific to the
details of the continuum model used. It is, in fact, a feature of models
involving hyperbolic conservation laws with forcing terms under rather
generic conditions. To illustrate this point, 
we shall briefly consider the equations presented by Aw \& Rascle
\cite{aw_rascle2000} and Helbing \cite{helbing2008b}. These models were
developed in response to the criticisms of Daganzo \cite{daganzo1995},
who argued that second order models are necessarily flawed --- they
predict, for example, shocks overtaking individual vehicles in unsteady
traffic flow, and negative vehicle speeds at the end of a stopped queue.

Aw \& Rascle \cite{aw_rascle2000} overcome such limitations by applying
a convective, rather than a spatial, derivative when modeling the effects
of preventive driving (the anticipation term in their nomenclature). This
leads to the momentum equation (\ref{eq:PWmom1}) being replaced by
\beq \label{eq:ARmom1}
  (u+p)_t+u\,(u+p)_x=\frac{1}{\tau}\,\left(\tu-u\right)\/.
\eeq
By substitution of the mass continuity equation (\ref{eq:PWmass1}),
(\ref{eq:ARmom1}) can be re-written as
\beq \label{eq:ARmom2}
  u_t+(u-c^2\,\rho)\,u_x=\frac{1}{\tau}\,\left(\tu-u\right)\/.
\eeq
Introducing the self-similar variable $\eta\/$ defined by
(\ref{eq:eta}), it can be shown that
\beq \label{eq:ARmom3}
  \frac{\dd\/u}{\dd\/\eta} =
  \frac{(u-s)\,(\tu-u)}{(u-s)^2-m\,c^2}\/,
\eeq
which is identical to (\ref{eq:PWmom2}), except that the sonic point is
now predicted to occur when $u-s=m^{1/2}c\/$. As before, the singularity
of (\ref{eq:ARmom3}) --- i.e.~the sonic point, where the denominator
vanishes --- is regularized by aligning the sonic point with a root of
$\tu-u\/$. The remainder of the analysis is entirely similar to that
outlined previously for (\ref{eq:PWmass1} -- \ref{eq:PWmom1}), with
appropriate modifications to the shock conditions. Under the Aw \& Rascle
formulation, the Rankine-Hugoniot condition corresponding to the
conservation of momentum takes the form
\beq \label{eq:RH3}
  s\,[\rho\,(u+p)] = [\rho\,u\,p+\rho\,u^2]\/,
\eeq
which replaces (\ref{eq:RH2}).

Helbing \cite{helbing2008b} generalized Aw \& Rascle's model one step
further by defining two traffic pressures, both functions of $\rho\/$
and $u\/$, such that the momentum equation reads
\beq \label{eq:helbing1}
  u_t+u\,u_x+\frac{1}{\rho}\,\frac{\partial\/p_1}{\partial\/\rho}\,\rho_x
  +\frac{1}{\rho}\,\frac{\partial\/p_2}{\partial\/u}\,u_x
  =\frac{1}{\tau}\,(\tu-u)\/.
\eeq
Proceeding as above, the following familiar expression can be readily
obtained:
\beq \label{eq:helbing2}
 \frac{\dd\/u}{\dd\/\eta} =
 \frac{(u-s)\,(\tu-u)}{(u-s)^2-\varsigma^2}\/,
\eeq
where
\beq \label{eq:varsigma}
 \varsigma^2 = c_1^2+m\,c_2^2\, ,
\eeq
in which
\bmath
  c_1^2 =   \frac{\partial\/p_1}{\partial\/\rho}
  \quad \mbox{and} \quad
  c_2^2 = - \frac{1}{\rho^2}\,\frac{\partial\/p_2}{\partial\/u}\/.
\emath
The Payne-Whitham and Aw \& Rascle results may be recovered from
(\ref{eq:helbing2}) by setting, respectively, $p_1=p\/$, $p_2=0\/$; and
$p_1=0\/$, $p_2=-u\,\rho^2\,\dd\/p/\dd\/\rho\/$.
%
%
\section{Phase plane analysis}
\label{sec:pp}
Generalizing the analyses of \S~\ref{sec:PW} and \S~\ref{sec:AR},
$\dd\/u/\dd\/\eta\/$ may be expressed as
\beq  \label{eq:genmom1}
  \frac{\dd\/u}{\dd\/\eta}=
  \frac{(u-s)\,(\tu-u)}{(u-s)^2-\Omega\,c^2}\/,
\eeq
where $\Omega=1\/$ and $m\/$, respectively, for the Payne-Whitham and
Aw \& Rascle models. Further information regarding model behavior near
the sonic point may be gleaned by introducing the phase plane variable
$\xi\/$, and rewriting (\ref{eq:genmom1}) as the following pair of
ordinary differential equations
\begin{eqnarray}  
  \frac{\dd\/u}{\dd\/\xi} & = & (u-s)\,(\tu-u)=F_1(u)\/, \label{eq:pp1}\\
  \frac{\dd\/\eta}{\dd\/\xi} & = &
   (u-s)^2-\Omega\,c^2=F_2(u)\/.\label{eq:pp2}
\end{eqnarray}
Note that the sonic point is a critical point of (\ref{eq:pp1}) and
(\ref{eq:pp2}). The Jacobian, $\mathcal{J}\/$, of the above pair of
equations is then given by
\beq \label{eq:mathcalJ}
\mathcal{J}=\left[
\begin{array}{cc}
            F_{1,u} & 0 \\
            F_{2,u} & 0
\end{array}
\right]\/,
\eeq
where
\begin{eqnarray}
  F_{1,u} & = & -(u-s)\left\{\frac{m\,\tu_{\rho}}{(u-s)^2}+1\right\}
  +\tu - u\/, \label{eq:F1u}\\
  F_{2,u} & = & 2\,(u-s)+\frac{\Omega\,m}{(u-s)^2}\,p_{\rho \rho}\/,
  \label{eq:F2u}
\end{eqnarray}
in which
\bmath
  \tu_{\rho} \equiv
  \frac{\dd\/\tu}{\dd\/\rho} < 0
  \quad \mbox{and} \quad
  p_{\rho \rho} \equiv \frac{\dd^2\/p}{\dd\/\rho^2} =
  \frac{\dd\/c^2}{\dd\/\rho}\/.
\emath
Therefore, at the sonic point the eigenvalues ($r_1\/$ and $r_2\/$)  of
$\mathcal{J}\/$ are given by
\beq \label{eq:eva}
  r_1=0 \quad \mbox{and} \quad
  r_2=\frac{m\,|\tu_{\rho}|}{\Omega^{1/2}\,c}-\Omega^{1/2}\,c
     =F_{1,u}\/.
\eeq
When $r_2 > 0\/$ (respectively $r_2 < 0\/$), $\,\dd\/u/\dd\/\eta > 0\/$
(respectively $\dd\/u/\dd\/\eta < 0\/$) at the sonic point. Because of the Lax
entropy conditions described earlier, $u\/$ should
%
%
be a monotonically increasing function of $\eta\/$ away
from a shock, which in turn requires
\beq
  r_2 > 0 \quad \iff \quad m\,|\tu_{\rho}| > \Omega \,c^2\/.
\eeq
Unfortunately, since $r_1 = 0\/$, the critical point is linearly
degenerate. Thus a complete analysis of the solution behavior near this
critical point requires a careful, but ultimately tangential, examination
of the leading order contributions by nonlinearities. Nevertheless, an
interesting observation can be made:
\beq \label{eqn:rrr-4.8}
  \left. \parbox{4.0in}{
     \emph{As we illustrate by way of example in \S~\ref{sec:example},
     $r_2=0\/$ coincides with the boundary wherein a constant uniform
     base state becomes unstable to infinitesimal disturbances} --- see
     figure \ref{fg:regime1} and Appendix~\ref{sec:appC}.}\;\; \right.
\eeq
The theoretical underpinnings of this coincidence are not entirely clear.
However, our numerical experiments show that there is a strong connection
between jamitons and instabilities:
\beq \label{eqn:rrr-4.9}
  \left. \parbox{4.0in}{
     \emph{When a uniform traffic state is linearly unstable, the
     instability consistently saturates into a jamiton-dominated
     state.}}\;\; \right.
\eeq
For the more general analysis of Helbing \cite{helbing2008b} discussed at
the end of \S~\ref{sec:AR}, $F_{2,u}\/$ is given by
\begin{eqnarray}
  F_{2,u} & = & 2\,(u-s)+\frac{m}{(u-s)^2}\,\varsigma^2_\rho\/,
  \nonumber \\
          & = & 2\,(u-s)+\frac{m}{(u-s)^2}\,p_{1,\rho \rho}+
                 \frac{2(u-s)}{m}\,p_{2,u}-p_{2,u \rho}\/, 
\label{eq:F2u_helbing}
\end{eqnarray}
where $\varsigma^2\/$ is defined by (\ref{eq:varsigma}). Given the form
of the Jacobian matrix $\mathcal{J}\/$, modifying $F_{2,u}\/$ does not
alter the eigenvalues $r_1\/$ and $r_2\/$.
%
%
\section{An example} \label{sec:example}
%
\subsection{Preliminaries} \label{sec:prelim}
To make the ideas of the previous sections more concrete, we consider
herein particular forms for $p\/$ and $\tu\/$, and carefully examine the
resulting range of solutions. As alluded to above, various expressions
for $p\/$ and $\tu\/$ have been proposed in the traffic literature.
Consistent with the spirit of previous studies (e.g.~%
\cite{kerner_konhauser1993, kerner_klenov_konhauser1997}), our motivation
is to select relatively simple functions so that the concepts of
\S~\ref{sec:PW} are succinctly illustrated with minimum algebraic
investment. Thus, a Lighthill-Whitham-Richards forcing term of the form
\beq \label{eq:LWR}
 \tu=\tu_0\,\left(1-\frac{\rho}{\rho_M}\right)\/,
\eeq
is chosen. Moreover, by analogy with the shallow water equations
\cite{dressler1949}, we select
\beq \label{eq:pres1}
 p=\mbox{$\frac{1}{2}$}\,\beta\,\rho^2\/,
\eeq
so that $p_{\rho}=c^2 \propto \rho\;$~\cite{aw_rascle2000}. Alternatively,
one could define $p\/$ to be singular in $\rho\/$, such that
$p_\rho \propto \rho^{\varphi_1}/(\rho_M-\rho)^{\varphi_2}\/$ for some
$\varphi_1\/,\,\varphi_2>0\/$. Whereas this, more complicated, choice for
$p\/$ enforces $\rho_M\/$ as the maximum traffic density, the resulting
algebraic relations become somewhat unwieldy. We therefore defer
consideration of singular pressure functions to future studies.

Applying the above definitions to the Payne-Whitham model of
\S~\ref{sec:ipf} yields
\beq \label{eq:PWmom3}
 \frac{\dd\/u}{\dd\/\eta} =
 \frac{(u-s)\/\left\{\tu_0\/
 \left(1-\frac{m}{\rho_M\,(u-s)}\right)-u\right\}}{%
 (u-s)^2-\frac{\beta\,m}{(u-s)}}\/.
\eeq
The sonic point is then defined by
\beq \label{se:sonic1}
 u-s=(\beta\,m)^{1/3}\/,
\eeq
while the zeros of the numerator are given by
\beq \label{eq:zeros1}
  u = \{u_1\/,\,u_2\}
    = s + \mbox{$\frac{1}{2}$}\,(\tu_0-s)
     \pm \mbox{$\frac{1}{2}$}(\tu_0-s)\,
     \left\{1-\frac{4\,\tu_0\,m}{\rho_M\,(\tu_0-s)^2}
     \right\}^{1/2}\/.
\eeq
The regularization condition (\ref{eq:sonic0}) may then be
written as
\beq \label{eq:reg1}
  u_2 = s+(\beta\,m)^{1/3}\/.
\eeq
Equations (\ref{eq:zeros1}) and (\ref{eq:reg1}) yield a cubic polynomial
in $s\/$, with at most two physically-relevant roots. With $p\/$
defined by (\ref{eq:pres1}), the Rankine-Hugoniot conditions specified
in (\ref{eq:RH1}) and (\ref{eq:RH2}) may be combined to yield
\beq \label{eq:dressler}
  \frac{u^--s}{u^+-s} = \frac{\rho^+}{\rho^-} =
  \mbox{$\frac{1}{2}$}\left\{-1+(1+8\,M_-^2)^{1/2}\right\}\/,
\eeq
where $M_-=|u^--s|/(\beta\,\rho^-)^{1/2}\,$ is the upstream Mach
number \cite{whitham1974, leveque1992}.
%
%
Finally, the non-trivial eigenvalue, $r_2\/$, of the Jacobian matrix
given by (\ref{eq:mathcalJ}), is, at the sonic point,
\beq \label{eq:r2}
  r_2 = \frac{\tu_0}{\rho_M}\left(\frac{m^2}{\beta}\right)^{1/3}
      - \left(\beta\,m\right)^{1/3}\/.
\eeq

Having regularized the ordinary differential equation given by
(\ref{eq:PWmom3}) --- i.e.~upon defining $u_2\/$ by (\ref{eq:reg1}) and
canceling $u-u_2\/$ from the numerator and denominator, the resultant
differential equation has an exact, albeit implicit, solution given by
\begin{eqnarray}
  \eta & = & u^+ - u + \frac{(\beta\,m)^{2/3}}{u_1-s}\,
            \mathrm{ln}\left(\frac{u-s}{u^+-s}\right) \nonumber \\
       &   & + \left\{(\beta\,m)^{1/3} + u_1 - s +
             \frac{(\beta\,m)^{2/3}}{u_1-s}\right\}\,
             \mathrm{ln}\left(\frac{u_1-u^+}{u_1-u}\right)\/.
\label{eq:exact1}
\end{eqnarray}
A similar result is obtained in the study of roll waves in shallow water
systems -- see e.g.~(4.18) in \cite{dressler1949}.

By definition, $u=u^-\/$ when $\eta=\lambda\/$, where $\lambda\/$ is the
length of the periodic roadway. Therefore
\begin{eqnarray}
  \lambda & = & u^+ - u^- + \frac{(\beta\,m)^{2/3}}{u_1-s}\,
                \mathrm{ln}\left(\frac{u^--s}{u^+-s}\right) \nonumber \\
          &   & + \left\{(\beta\,m)^{1/3} + u_1 - s +
                  \frac{(\beta\,m)^{2/3}}{u_1-s}\right\}\,
                  \mathrm{ln}\left(\frac{u_1-u^+}{u_1-u^-}\right)\/.
\label{eq:lambda1}
\end{eqnarray}
Starting from (\ref{eq:mathcalN1}), the total number of vehicles along
the periodic circuit can be computed from
\beq \label{eq:mathcalN2}
  \mathcal{N }= \int_0^{\lambda} \rho\,\dd\/x
  = \int_0^{\eta_{max}} \rho\,\dd\/\eta
  =m\, \int_{u^+}^{u^-}\frac{1}{u-s}\,\left(
   \frac{\dd\/u}{\dd\/\eta}\right)^{-1}\dd\/u\/.
\eeq
Application of (\ref{eq:PWmom3}) in (\ref{eq:mathcalN2}) yields then the
explicit result
\begin{eqnarray}
  \mathcal{N} & = & m\,\tau\,\left\{\frac{(\beta\,m)^{2/3}}{u_1-s}
     \left\{\frac{u^--u^+}{(u^--s)\,(u^+-s)}\right\} +
     \right. \nonumber \\
  & & \frac{(\beta\,m)^{1/3}}{u_1-s} \left\{ \frac{(\beta\,m)^{1/3}}{u_1-s}
      + 1 \right\}\,
      \mathrm{ln}\left\{
      \frac{(u^--s)\,(u_1-u^+)}{(u^+-s)\,(u_1-u^-)}\right\}
     \nonumber \\
  & & + \left. \mathrm{ln}\left(\frac{u_1-u^+}{u_1-u^-}\right)\right\}\/.
\label{eq:mathcalN3}
\end{eqnarray}

Comparable exact solutions may also be determined when $p \propto \rho\;$,
rather than $p \propto \rho^2\/$. These are presented in Appendix \ref{sec:appA}.
%
\subsection{Numerical method}
\label{sec:num}
In order to validate the aforementioned theoretical solutions
and assess jamiton stability, we performed numerical
simulations using a Lagrangian particle method \cite{monaghan1988}. In
this method, each discrete particle, $i\/$, is assigned an initial
position, $x_i\/$, and speed, $u_i\/$, which subsequently changes in
time according to  (\ref{eq:cases} -- \ref{eq:particle_method_rhox}). The
mass balance equation (\ref{eq:PWmass1}) is satisfied identically as
the particles move, i.e.~the numerical scheme is mass-conservative by
construction.
Importantly, the number of particles is typically two orders of magnitude
larger than the number of vehicles. Thus, although a Lagrangian approach
is employed, the numerical scheme constitutes a macroscopic, not a
microscopic, description of traffic flow, albeit one with an intuitive
link between the particle and vehicle density. 

In general terms, the numerical method solves the differential equations
\beq
\begin{cases}
  \dot x_i &= u_i\/, \\ \dot u_i &= a_i\/,
\end{cases}
\label{eq:cases}
\eeq
where the particle acceleration, $\dot u_i\/$, is expressed as
\beq
 \frac{\dd\/u}{\dd\/t}(x_i\/,\,t) =
 u_t(x_i\/,\,t)+u\,u_x(x_i\/,\,t)\/,
\eeq
and
\beq
  a_i = - \frac{c(\rho(x_i\/,\,t))^2}{\rho(x_i\/,\,t)}\,
  \rho_x(x_i\/,\,t) + \frac{1}{\tau}\,\left\{
  \tu(\rho(x_i\/,\,t))-u(x_i\/,\,t)\right\}\/.
\eeq
The distance between adjacent particles $i\/$ and $i+1\/$ is defined by
$d_{i+\frac{1}{2}} = x_{i+1}-x_i\/$. Then the inter-particle density is
computed by
\beq \label{eq:partdens1}
  \rho_{i+\frac{1}{2}} = \frac{\vartheta}{d_{i+\frac{1}{2}}}\,,
\eeq
where $\vartheta=\mathcal{N}/N_p\/$ --- in which $\mathcal{N}\/$ is the
number of vehicles (as specified by (\ref{eq:mathcalN1})) and $N_p\/$ is
the number of particles.
%
%
From (\ref{eq:partdens1}), we define the vehicle density and the density
gradient using a non-equidistant finite-difference stencil
\cite{dilts1999}: 
\beq \label{eq:partdens2}
  \rho_i = \frac{d_{i+\frac{1}{2}}\,\rho_{i-\frac{1}{2}} +
                 d_{i-\frac{1}{2}}\,\rho_{i+\frac{1}{2}}}
                {d_{i+\frac{1}{2}} + d_{i-\frac{1}{2}}}\/,
\eeq
\beq \label{eq:particle_method_rhox}
  \rho_{i,x} = \frac{\rho_{i+\frac{1}{2}}-\rho_{i-\frac{1}{2}}}
              {\min\{d_{i+\frac{1}{2}}\/,\,2\,d_{i-\frac{1}{2}}\}+
               \min\{d_{i-\frac{1}{2}}\/,\,2\,d_{i+\frac{1}{2}}\}}\/.
\eeq
The denominator in (\ref{eq:particle_method_rhox}) is chosen so that, at
the location of the shock, a given particle is influenced only by its
nearest neighbor. The numerical scheme is stable --- however, it produces
bounded but unphysical oscillations wherever the vehicle density or
speed changes abruptly. We suppress these features by adding a small
amount of numerical viscosity, 
so that at each time step the velocity profile is smoothed out.
%
\subsection{Results}
\label{sec:res}
Here we compare the theory of self-sustained traffic jams developed in
\S~\ref{sec:PW} and \S~\ref{sec:prelim}, and numerical solutions of
(\ref{eq:PWmass1} -- \ref{eq:PWmom1}) --- obtained using the algorithm
of \S~\ref{sec:num}. Employing the forcing and traffic pressure terms
specified in (\ref{eq:LWR}) and (\ref{eq:pres1}), the equations have the
non-dimensional form
\begin{equation} \label{eqn:rrr-5.19}
 \left. \begin{array}{rcl}
    \rho^*_{t^*} + (u^*\,\rho^*)_{x^*} & = & 0\/,\\
    \Gamma_1\,\left( u^*_{t*} + u^*\,u^*_{x^*} \right) +
    (\Gamma_2/\rho^*)\,p^*_{x^*} & = &
    1 - \rho^* - u^*\/, \rule{0mm}{3ex}
 \end{array} \right\}
\end{equation}
where
$\,\Gamma_1 = \tau \,\tu_0 \rho_M\/$,
$\,\Gamma_2 = \beta\,\tau\rho_M^2/\tu_0\/$, and
$\,p^* = \frac{1}{2}(\rho^*)^2\/$.
The non-di\-men\-sional and dimensional variables are related via
\begin{equation} \label{eqn:rrr-5.20}
 \rho = \rho_M\,\rho^*\/, \quad u = \tu_0\,u^*\/, \quad
 x = \ell\,x^*\/, \quad \mbox{and} \quad
 t = (\ell/\tu_0)\,t^*\/,
\end{equation}
where $\ell = 1/\rho_M\/$ is the effective vehicle length, defined earlier
in \S~\ref{sec:ipf}, the non-dimensional traveling wave speed is
$s^* = s/\tu_0\/$, and the parameter $\tau$ is the adjustment time-scale
required for an individual vehicle to effect an $\mathcal{O}(1)\/$ change
in speed.

\begin{figure}
 \vspace*{-8.5cm} 
 \hspace*{-1cm}
 \includegraphics[width=15cm]{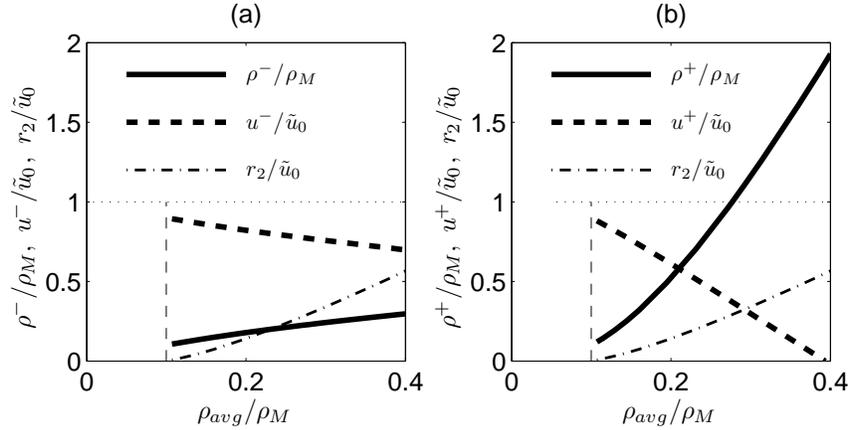}
 \noindent
 \vspace*{-7.25cm}
\caption{Parameters characterizing the exact jamiton solutions in
   \S~\ref{sec:prelim}, for a fixed road length $\lambda=500$\,m, as
   functions of the number of vehicles, $\mathcal{N}\/$. Specifically,
   curves of $\,\rho^-/\rho_M\/$, $\,\rho^+/\rho_M\/$, $\,u^-/\tu_0\/$,
   and $\,u^+/\tu_0\/$ versus
   $\rho_{avg}/\rho_M = \mathcal{N}/(\lambda\,\rho_M)\/$ are shown, in
   addition to $\,r_2/\tu_0\/$ --- as given in (\ref{eq:r2}). The
   vertical dashed line is the stability boundary specified by
   (\ref{eq:linstab1}). The equations and parameters are as in
   (\ref{eqn:rrr-5.19} -- \ref{eq:Gammas}). Note that {\it (i)}
   $r_2\/$ goes through $0\/$ precisely at the stability boundary ---
   see (\ref{eqn:rrr-4.8}); {\it (ii)} the peak density $\rho^+\/$
   reaches $\rho_M\/$ before $u^+\/$ becomes negative --- see Appendix
   \ref{sec:uplus0}.}
\label{fg:regime1}
\end{figure}

We consider as canonical modeling parameters $\ell = 5\,$m,
$\,\tu_0 = 30\,$m/s, $\,\tau = 3.\bar{3}\,$s, and
$\beta = 450\,$m$^3$/s$^2\/$, whereby
\beq \label{eq:Gammas}
 \Gamma_1 = 20 \quad \mbox{and} \quad \Gamma_2 = 2\/.
\eeq
%
%
Note that $\beta=c^2/\rho$ is related to the speed at which disturbances propagate through traffic. Because a description, both theoretical and numerical, of jamitons is the
central focus of the present analysis, we deliberately select a value
for $\beta\/$ towards the lower end of its representative range, so as
to facilitate an explicit display of jamiton properties and behavior.
Hence, $\beta\/$, whose exact numerical value is likely difficult to estimate in any event, 
is chosen so that infinitesimal perturbations to a
uniform base state develop into jamitons when the base-state density
is 10\% of $\rho_M\/$ --- see Appendix \ref{sec:appC} and, more
particularly, the stability condition (\ref{eq:linstab1}). Whereas
larger numerical values for $\beta\/$ could be selected, corresponding
to a more restrictive instability condition, the jamitons observed in
these cases, though qualitatively equivalent to those described below,
are somewhat less easy to visualize.
The practical ramifications associated with choosing a larger value for
$\beta\/$ are briefly addressed
%
%
%
when we discuss (after examining the results in
figure \ref{fg:regime1}) the possibility of vehicular collisions.

%
\begin{figure}
 \vspace*{-8.5cm} 
 \hspace*{-1cm}
 \includegraphics[width=15cm]{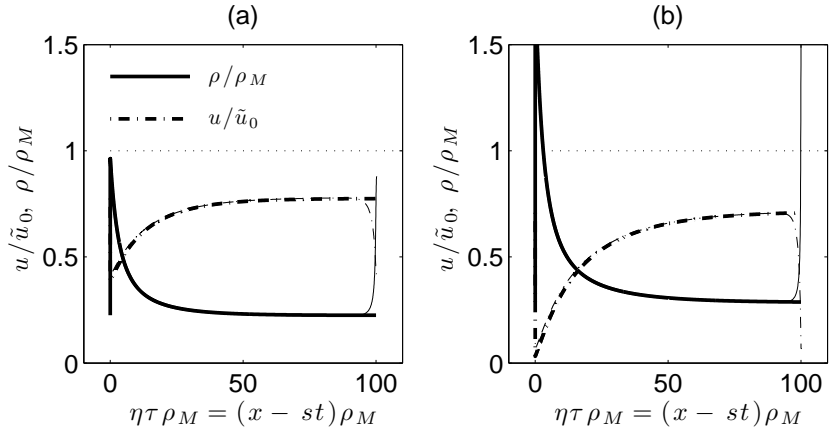}
 \noindent
 \vspace*{-7cm}
\caption{Comparison of theoretical (thick curves) and numerical (thin
   curves) solutions. The numerical solutions are the final, asymptotic
   state of an evolution started with a small perturbation of a uniform
   unstable base state --- see (\ref{eq:linstab1}). The agreement of this
   final state with the jamiton corresponding to the same road length,
   $\lambda\/$, and number of vehicles, $\mathcal{N}\/$, is remarkable.
   The plots show profiles of $\rho/\rho_M\/$ and $u/\tu_0\/$ versus
   $\eta\,\tau\,\rho_M=(x-st)\,\rho_M\/$. The equations and parameters
   are as in (\ref{eqn:rrr-5.19} -- \ref{eq:Gammas}), with
   $\lambda = 100\,\ell = 500\,$m. In panel (a) $\,\mathcal{N}=27\/$
   ($\rho_{avg}/\rho_M=0.272\/$). In panel (b) $\,\mathcal{N}=38\/$
   ($\rho_{avg}/\rho_M=0.384\/$).}
\label{fg:urhoeta}
\end{figure}

The steady-state variations of $u^*\/$ and $\rho^*\/$, as functions of the
non-di\-men\-sional variable $\eta\,\tau\,\rho_M = x^* - s^*\,t^*\/$, are
shown in figure \ref{fg:urhoeta} for a circular road of length
$\lambda = 100\,\ell\/$, with two different choices for the conserved
number of vehicles, $\mathcal{N}\/$. The shock occurs at the two extreme
ends of the horizontal domain, and connects the ratios
$u^-/\tu_0\/$ to $u^+/\tu_0\/$, and $\rho^-/\rho_M\/$ to
$\rho^+/\rho_M\/$. In figure \ref{fg:urhoeta}\,a, $\mathcal{N}=27\/$, and
the maximum traffic density (i.e.~$\rho^+\/$) is predicted to be just
below $\rho_M\/$. 
Conversely, in figure \ref{fg:urhoeta}\,b, $\,\mathcal{N}\/$ is increased
to $38\/$, which results in $\rho^+>\rho_M\/$. The physical implications
of this result are considered in the following two paragraphs. 
Both theoretical and numerical data are
included in figure \ref{fg:urhoeta}. The comparison is very favorable,
except right at the shock location --- where the numerical scheme smears
the shock.

Figure \ref{fg:regime1} indicates, as a function of the normalized average
traffic density $\rho_{avg}/\rho_M\/$ (where
$\rho_{avg}=\mathcal{N}/\lambda\/$), the range of possible solutions
allowed by the model equations in \S~\ref{sec:prelim}. The thick solid
curves of figures \ref{fg:regime1}\,a,b show $\rho^-/\rho_M\/$ and
$\rho^+/\rho_M\/$, respectively, whereas the thick dashed curves show
$u^-/\tu_0\/$ and $u^+/\tu_0\/$, respectively. As with the discussion
in \S~\ref{sec:PW}, we observe that, for a prescribed road
length $\lambda\/$ and model parameters $\rho_M\/$, $\tu_0\/$,
$\beta\/$, and $\tau\/$, the average density $\rho_{avg}\/$
(hence $\mathcal{N}\/$) uniquely determines the flow conditions to either
side of the shock. Although the solid and dashed curves of
figure~\ref{fg:regime1}\,a do not proffer any particularly meaningful
insights, those of figure~\ref{fg:regime1}\,b are significant in that
they predict the following forms of model breakdown: $\rho^+>\rho_M\/$
when $\rho_{avg}/\rho_M > 0.277\/$ and $u^+<0\/$ when
$\rho_{avg}/\rho_M>0.391\/$.\footnote{In Appendix \ref{sec:uplus0}, we
   verify that $\rho \to \rho_M\/$ before $u^+ \to 0\/$.}

Rather than signifying a fundamental modeling flaw, observations of model
breakdown provide helpful guidance in the safe design of modern roadways.
In particular, for the choice of parameters germane to
figure~\ref{fg:regime1}, vehicular collisions (i.e.~$\rho \to \rho_M\/$)
are anticipated once the average center-to-center separation between
adjacent vehicles falls below approximately $3.6\,\ell\/$. 
Needless to say, this simple result is not universal for all types of
traffic flow or roadway conditions. We consider herein a periodic track
of prescribed length (500\,m), particular functional forms for the
traffic pressure, $p\/$, and equilibrium speed, $\tu\/$, and a liberal
numerical value for $\beta\/$ such that jamitons appear even in
relatively light traffic. In particular, choosing a larger value for
$\beta\/$ would delay, though not necessarily avoid, the onset of
vehicular collisions. 

The roadway's carrying capacity could, in principle, be increased if two
or more traveling waves were to be forced rather than the single density
spike considered here. However, our numerical simulations suggest that
such bi- or multi-modal structures are often unstable, and that they
quickly coalesce into a single structure. By
contrast, unimodal traveling wave solutions of the type illustrated in
figure~\ref{fg:urhoeta} appear to be stable to small perturbations ---
except in the special case where the periodic roadway is made to be quite
long, i.e.~$\gtrsim 2\,$km for the parameters appropriate to
figures~\ref{fg:urhoeta} and \ref{fg:regime1}. On an extended circuit,
the traveling wave solutions have a long, nearly constant state
downstream of the shock --- whose average density exceeds the requirement
for linear stability discussed below. Thus infinitesimal perturbations
may grow, leading to further traveling waves, albeit of slightly
different amplitude from the original. Owing to the circuit length
(i.e.~$\gtrsim 2\,$km, as compared to the $500\,$m long circuit
considered in figure \ref{fg:urhoeta}), these traveling waves are
dynamically independent in that wave coalescence does not occur for a
rather long time, if at all.

%
%

While the above observations would benefit from the development of a more
fundamental framework, it can be difficult to describe analytically the
spatio-temporal stability of self-sustained traveling waves, due primarily
to the transonic nature of the solution (see
e.g.~Stewart \& Kasimov \cite{StewartKasimovSIAP05},
Balmforth \& Mandre \cite{balmforth_mandre2004} and
Yu \& Kevorkian \cite{yu_kevorkian1992}). Work is on-going to adapt
nonlinear stability analyses for specific application to continuum
traffic models. 

In Appendix \ref{sec:appC} we compute the boundary between linearly
stable and unstable uniform base states, which is given by
\beq \label{eq:linstab1}
\frac{\rho}{\rho_M}=\frac{\beta\,\rho_M}{\tu_0^2}\/.
\eeq
In connection with this boundary, we point out the following facts, which
together with (\ref{eqn:rrr-4.8} -- \ref{eqn:rrr-4.9}), reinforce the
point that there is a strong connection between jamitons and
uniform flow instabilities.
\begin{itemize}
 %
 \item
 Figure \ref{fg:regime1} shows that $[\rho\,] \to 0\/$ and $[u] \to 0\/$,
 as $\rho_{avg}/\rho_M \downarrow \beta\,\rho_M/\tu_0^2\/$. Thus
 the jamiton amplitude becomes smaller as the corresponding
 uniform flow
 %
 %
 becomes less unstable, and vanishes altogether in the limit.
 %
 \item
 The stability criterion (\ref{eq:linstab1}) also specifies the
 location at which $u_1=u_2\/$, where $u_1\/$ and $u_2\/$ are defined in
 figure \ref{fg:cartoon1}. Once $u_1\/$ and $u_2\/$ coalesce, no
 self-sustained shock wave may occur --- since condition (a) in
 (\ref{eqn:rrr-2.11}) fails. This is consistent with the discussion of the previous bullet, of course.
\end{itemize}      
Thus jamitons become possible when the corresponding uniform state
becomes unstable
and --- see  (\ref{eqn:rrr-4.9}) --- the basic flow state changes from
uniform flow to a jamiton-dominated state. In other words, evidence
indicates that
%
%
\begin{equation} \label{eqn:rrr-5.BIF}
  \left.
  \parbox{4.0in}{
     \emph{A crucial bifurcation in the traffic flow behavior occurs
           at the stability boundary prescribed by (\ref{eq:linstab1}).}}
  \right.
\end{equation}
We defer a more in-depth investigation of this question for future work.

Finally, note that although the boundary specified by (\ref{eq:linstab1})
is mathematically robust, in practice it may become ``fuzzy'' owing to
the possible breakdowns of the continuum hypothesis, especially at low
vehicle concentrations.
%
%
\section{Conclusions}
\label{sec:conc}
%
In this work, we have found and successfully exploited a strong similarity
between gas-dynamical detonation waves and shocks in traffic flow, in
order to develop a theory of ``jamitons,'' steady self-sustained traffic
shocks.
%
Jamitons naturally arise from small instabilities in relatively dense
traffic flow, and can be interpreted as saturated phantom jams.
While a single jamiton may not necessarily significantly delay individual
vehicles, a succession of jamitons, as might arise during rush hour, for
example, is expected to frustrate motorists over long lengths of roadway.
Moreover, jamitons represent regions in which the traffic density
increases dramatically over a relatively short distance
\cite{sugiyama_fukui_kikuchi_hasebe_nakayama_nishinari_tadaki_yukawa2008}
and, as such, are hot spots for vehicular collisions. 

The analogy drawn above goes beyond the weaker analogy with {inert}
shock waves in gas dynamics considered by Kerner, Klenov \&
Konh{\" a}user \cite{kerner_klenov_konhauser1997}. Unlike inert shocks,
detonation waves can be self-sustained, due to the existence of a sonic
point in direct correspondence with the traffic model we consider here.  

Using the Lax entropy conditions for hyperbolic conservation laws, we show
that for realistic pressure and equilibrium-speed functions, the only
allowable self-sustained shocks are those that are overtaken by individual
vehicles. Moreover, for the simple, but widely-applied, choices
$\tu=\tu_0(1-\rho/\rho_M)\/$ and $p \propto \rho^\gamma\/$, with
$\gamma=1\/,\,2\/$, we are able to describe the jamiton structure
analytically. Theoretical solutions show excellent agreement with the
output from direct numerical simulations of the governing system.
Examples of jamitons with $\gamma=2\/$ are illustrated in figure
\ref{fg:urhoeta}. Beyond a critical average density, we predict solutions
for which the density immediately downstream of the shock exceeds the
maximum allowable density, $\rho_M\/$, corresponding, physically, to a
state of vehicular collisions (figure \ref{fg:regime1}). Such instances
offer important design insights for roadways; most obviously, it is
advantageous to choose speed limits and roadway carrying capacities
so as to avoid circumstances where densities with $\rho \gtrsim \rho_M\/$
are ``triggered'' (say, by a jamiton) anywhere within the domain.

Having identified self-sustained traveling wave solutions in traffic flow,
a major objective of future research is to ascertain their spatio-temporal
stability. Some of the analytical challenges associated with this line of
inquiry are identified in \S~\ref{sec:res}. Resolving issues of stability
offers the possibility of increasing roadway efficiency, for example, by
exciting multiple traveling waves of relatively low density as compared to
the single density spike exhibited in figure \ref{fg:urhoeta}. We hope to
report on the results of such an analysis soon.  

\vspace*{0.5cm}
\noindent \emph{Acknowledgments} \newline \noindent
Partial support for J.-C.~Nave, R.~R.~Rosales and B.~Seibold was
provided through NSF grant DMS-0813648.
Funding for A.~R.~Kasimov was provided through the AFOSR Young
Investigator Program grant FA9550-08-1-0035
(Program Manager Dr.~Fariba Fahroo).
We thank Dr.~P.M.~Reis for bringing to our attention the study of
Sugiyama \emph{et al.}
\cite{sugiyama_fukui_kikuchi_hasebe_nakayama_nishinari_tadaki_yukawa2008}.
%
%
\appendix
\section{Exact solutions when $p \propto \rho\/$}
\label{sec:appA}
%
Equations (\ref{eq:exact1}), (\ref{eq:lambda1}) and (\ref{eq:mathcalN3})
specify, respectively, exact analytical solutions for $u\/$, $\lambda\/$
and $\mathcal{N}\/$ when $p \propto \rho^2\/$. Corresponding expressions,
valid when $p \propto \rho\/$, are given by
\beq \label{eq:exact2}
  \eta = u^+ - u + \left(u_1-s+\beta^{1/2}\right)\,
  \mathrm{ln}\left(\frac{u_1-u^+}{u_1-u}\right)\/,
\eeq
\beq \label{eq:lambda2}
  \lambda = u^+ - u^- + \left(u_1-s+\beta^{1/2}\right)\,
  \mathrm{ln}\left(\frac{u_1-u^+}{u_1-u^-}\right)\/,
\eeq
and
\beq \label{eq:mathcalN4}
  \mathcal{N} = m\,\tau\,\left\{\frac{\beta}{u_1-s}\,
  \mathrm{ln}\left(\frac{u^--s}{u^+-s}\right) +
  \left(1+\frac{\beta}{u_1-s}\right)\,
  \mathrm{ln}\left(\frac{u_1-u^+}{u_1-u^-}\right)\right\}\/,
\eeq
respectively. As before, the traffic speed is given as an implicit
function of $\eta\/$ in (\ref{eq:exact2}).
%
\section{Model behavior for $\rho>\rho_M\/$}
\label{sec:uplus0}
%
From the discussion of \S~\ref{sec:example}, vehicular collisions are
forecast once $\rho^+=\rho_M\/$, whereas nonsensical vehicle speeds
are predicted once $u^+<0\/$. It is demonstrated herein that the former
condition is necessarily achieved before the latter. 

Equation (\ref{eq:PWmass2}) shows that
\beq \label{eq:breakdown1}
 \rho^+\,(u^+-s) = \rho_M\,(u_M-s) = m>0\/,
\eeq
where $u_M\/$, which must be positive for jamitons to exist, is defined
in figure \ref{fg:cartoon1}\,b. As $u^+ \to 0\/$, the left-hand side of
(\ref{eq:breakdown1}) approaches $-\rho^+\,s\/$, demonstrating that
$s<0\/$. In this limit, therefore,
\beq \label{eq:breakdown2}
  \frac{\rho^+}{\rho_M} = \frac{u_M-s}{|s|}>1\/.
\eeq
This result can be understood intuitively by examining the functional
form of the Lighthill-Whitham-Richards forcing term (see (\ref{eq:LWR})):
$\tu<0\/$ requires $\rho>\rho_M\/$. Clearly, the sensible
alternative is to define $\tu=0\/$ for $\rho>\rho_M\/$. The point
is moot, however: this amounts to correcting the model equations in a
regime that is already physically unrealistic.
%
\section{Linear stability of the Payne-Whitham \\
         model considered in \S~\ref{sec:example}}
\label{sec:appC}
%
For a right-hand side forcing function of type (\ref{eq:LWR}), the
constant base state solution to (\ref{eq:PWmass1}) and (\ref{eq:PWmom1})
is given by
\beq \label{eq:basestate}
  \rho=\tilde{\rho}\/, \qquad
  u = \tu_0\,\left(1-\frac{\tilde{\rho}}{\rho_M}\right)\/.
\eeq
Following ideas discussed in Kerner \& Konh{\" a}user
\cite{kerner_konhauser1993}, Helbing \cite{helbing2008b} and elsewhere,
the linear stability of this base state can be explored by introducing
perturbation (hatted) quantities, defined such that
\beq \label{eq:pert1}
  \rho = \tilde{\rho}+\hat{\rho}\/, \qquad
     u = \tu_0\,\left(1-\frac{\tilde{\rho}}{\rho_M}\right)+\hat{u}\/,
\eeq
where $\hat{\rho}\/$ and $\hat{u}\/$ are expressed in terms of normal
modes by
\beq \label{eq:nm1}
  \hat{\rho} = \hat{R}\,\mathrm{e}^{\mathrm{i}\/k\/x+\sigma\/t}
  \quad \mbox{and} \quad \hat{u} =
  \hat{U}\,\mathrm{e}^{\mathrm{i}\/k\/x+\sigma\/t}\/.
\eeq
Here $k\/$ is the horizontal wave number and $\sigma\/$ is the
corresponding growth rate. Application of (\ref{eq:pert1}) and
(\ref{eq:nm1}) into (\ref{eq:PWmass1}) and (\ref{eq:PWmom1}) shows that
\beq
  \left[ \begin{array}{cc}
            \sigma+\mathrm{i}\,k\,\psi & \mathrm{i}\,k\,\tilde{\rho} \\
            {\displaystyle \frac{\tu_0}{\tau\/\rho_M}} +
            \mathrm{i}\,\beta\,k & \sigma + \mathrm{i}\,k\,\psi +
            \frac{1}{\tau} \rule{0mm}{3.7ex}
  \end{array} \right]
  \left[ \begin{array}{c}
            \hat{R} \\ \hat{U} \rule{0mm}{3.9ex}
  \end{array} \right]
  =
  \left[ \begin{array}{c}
            0 \\ 0  \rule{0mm}{4.3ex}
  \end{array} \right]\/,     \label{eq:linstab2}
\eeq
where, for notational economy, we have introduced
\beq \label{eq:psi}
  \psi = \tu_0\/\left(1-\frac{\tilde{\rho}}{\rho_M}\right)\/.
\eeq
Requiring that the determinant of the matrix from (\ref{eq:linstab2})
vanish shows that
\beq \label{eq:sigma1}
  \sigma = -\mathrm{i}\,k\,\psi-\frac{1}{2\,\tau}\,(1+\Upsilon)\/,
\eeq
in which
\beq \label{eq:bowtie1}
  \Upsilon^2 = 1 - 4\,k^2\,\beta\,\tilde{\rho}\,\tau^2 +
  4\,\mathrm{i}\,k\,\tu_0\,\tau\,\frac{\tilde{\rho}}{\rho_M}\/.
\eeq
Generically, $\Upsilon\/$ may be written as
$\Upsilon = \Lambda_1-\mathrm{i}\/\Lambda_2\/$, where
\beq \label{eq:LambdaA}
  \Lambda_1^2 - \Lambda_2^2 = 1 - 4\,k^2\,\beta\,\tilde{\rho}\,\tau^2\/,
\eeq
and
\beq \label{eq:LambdaB}
  \Lambda_1\,\Lambda_2 = -2\,k\,\tu_0\,\tau\,\frac{\tilde{\rho}}{\rho_M}\/.
\eeq
Eliminating $\Lambda_2\/$ from (\ref{eq:LambdaA}) and (\ref{eq:LambdaB})
yields the following polynomial expression:
\beq \label{eq:mathcalP}
  \mathcal{P}(\Lambda_1^2) = \Lambda_1^4 - (1 -
  4\,k^2\,\beta\,\tilde{\rho}\,\tau^2)\,\Lambda_1^2 -
  4\,k^2\,\tu_0^2\,\tau^2\,\frac{\tilde{\rho}^2}{\rho_M^2}=0\/.
\eeq
Linear stability requires a non-positive growth rate, i.e.
\beq \label{eq:linstab3}
  \mathrm{Real}(\sigma) \leq 0 \quad \Longleftrightarrow \quad
  0 \leq \Lambda_1^2 \leq 1    \quad \Longleftrightarrow \quad
  \mathcal{P}(1) \geq 0\/.
\eeq
From (\ref{eq:mathcalP}), the latter condition is satisfied provided
\beq \label{eq:linstab4}
  \beta \geq \frac{\tilde{\rho}\,\tu_0^2}{\rho_M^2}\/.
\eeq
This completes the derivation of (\ref{eq:linstab1}).
%
%

%
\bibliography{fluid_refs}
\bibliographystyle{unsrt}

\end{document}